\documentstyle[sprocl]{article}

\def\al{\alpha}
\def\be{\beta}
\def\ga{\gamma}
\def\de{\delta}

\def\ve{\varepsilon}

\def\et{\eta}
\def\th{\theta}

\def\la{\lambda}

\def\ps{\psi}

\def\Om{\Omega}
\def\mn{{\mu\nu}}

\def\prt{\partial}

\def\frac#1#2{{\textstyle{{#1}\over {#2}}}}
\def\half{{\textstyle{1\over 2}}}
\def\lsim{\mathrel{\rlap{\lower4pt\hbox{\hskip1pt$\sim$}}
    \raise1pt\hbox{$<$}}}
\def\gsim{\mathrel{\rlap{\lower4pt\hbox{\hskip1pt$\sim$}}
    \raise1pt\hbox{$>$}}}
\def\sqr#1#2{{\vcenter{\vbox{\hrule height.#2pt
         \hbox{\vrule width.#2pt height#1pt \kern#1pt
         \vrule width.#2pt}
         \hrule height.#2pt}}}}

\def\lrDmu{\stackrel{\leftrightarrow}{D_\mu}}
\def\lrDnu{\stackrel{\leftrightarrow}{D^\nu}}

\def\abs#1{\left|{#1}\right|}

\def\hg{ ^{199}{\rm Hg} }

\def\ber{ ^{9}{\rm Be}^+ }

\def\cs{ ^{133}{\rm Cs} }

\def\tmf{ \widetilde{m}_F }

\def\ol#1{\overline{#1}}

\newcommand{\beq}{\begin{equation}}
\newcommand{\eeq}{\end{equation}}
\newcommand{\bea}{\begin{eqnarray}}
\newcommand{\eea}{\end{eqnarray}}

\newcommand{\eq}[1]{Eq.\ (\ref{#1})}

\begin{document}

\title{ATOMIC PROBES OF NONCOMMUTATIVE FIELD THEORY}

\author{C. D. LANE}

\address{
 Department of Physics, 
 Berry College \\
 2277 Martha Berry Blvd., 
 Mount Berry, GA 30149-5004 \\
 Email: cdlane@mailaps.org}


\maketitle\abstracts{ 
 We consider the role of Lorentz symmetry in 
 noncommutative field theory. 
 We find that a Lorentz-violating 
 standard-model extension involving ordinary fields 
 is general enough to include 
 any realisitc noncommutative field theory as a subset. 
 This leads to various theoretical consequences, 
 as well as bounds 
 from existing experiments 
 at the level of (10 TeV)$^{-2}$
 on the scale of the noncommutativity parameter.
}


Following discovery that it could arise naturally in string theory, 
there has been a recent revival 
of the notion that spacetime may intrinsically involve 
noncommutative coordinates.\cite{cds} 
In noncommutative geometry, 
cartesian-like coordinates $\{x^\mu\}$ obey 
commutation relations
\beq
[x^\mu , x^\nu] = i\th^\mn
\quad ,
\label{coords}
\eeq
where the components of $\th^\mn$ are real 
and satisfy $\th^\mn=-\th^{\nu\mu}$. 
The nonzero $\th^\mn$ in \eq{coords} 
necessarily leads to violations of Lorentz symmetry.\cite{cpt99} 
In the present work, 
we primarily study the physical effects 
that arise from this Lorentz violation 
and consider the sensitivity of current experiments 
to possible realistic noncommutative field theories.\cite{chklo}

The following recipe may be used to generate 
a noncommutative quantum field theory: 
Begin with an ordinary theory, 
then replace all ordinary fields with noncommutative fields 
and all ordinary products with Moyal $\star$ products, 
defined 
by 
\beq
(f\star g)(x) := \exp{(\half i\th^\mn\prt_{x^\mu}\prt_{y^\nu})}
 f(x)g(y)|_{x=y}
\quad .
\label{moyal}
\eeq
In general, 
the resulting noncommutative theory 
will not support any ordinary gauge symmetry 
that the original ordinary theory supports. 
However, there may be a modified noncommutative 
gauge symmetry that is supported by the 
noncommutative field theory. 
In the present work, we concentrate on noncommutative 
quantum electrodynamics\cite{haya} (QED),
which has hermitian lagrangian
\beq
{\cal L} = \half i \ol{\widehat{\ps}} \star \ga^\mu 
 \stackrel{\leftrightarrow}{\widehat{D}_\mu} \widehat{\ps}
 - m \ol{\widehat{\ps}} \star \widehat{\ps} 
 - \frac{1}{4q^2} \widehat{F}_\mn \star \widehat{F}^\mn
\quad .
\label{nclagr1}
\eeq
Here, carets indicate noncommutative quantities, 
$\widehat{F}_\mn:=\prt_\mu\widehat{A}_\nu 
 -\prt_\nu\widehat{A}_\mu 
 -i[\widehat{A}_\mu,\widehat{A}_\nu]_\star$, 
$\widehat{D}_\mu\widehat{\ps}:=\prt_\mu\widehat{\ps} 
 -i\widehat{A}_\mu\star\widehat{\ps}$, 
and $\widehat{f}\star \stackrel{\leftrightarrow}{\widehat{D}_\mu} 
 \widehat{g}:=\widehat{f}\star\widehat{D}_\mu\widehat{g} 
 -\widehat{D}_\mu\widehat{f}\star\widehat{g}$.

Since the parameter $\th^\mn$ carries Lorentz indices, 
the application of Lorentz transformations to \eq{nclagr1} 
requires more care than usual. 
In particular, 
the two distinct types of Lorentz transformation,\cite{ck} 
observer and particle, 
must be distinguished. 
Observer Lorentz transformations 
leave the physics associated with \eq{nclagr1} unchanged, 
since the quadratic field operators and $\th^\mn$ 
transform as Lorentz tensors according to their Lorentz indices. 
Thus, \eq{nclagr1} is fully observer Lorentz symmetric. 
In contrast, 
particle Lorentz transformations 
treat the quadratic field operators as Lorentz tensors 
according to their Lorentz indices, 
but treat $\th^\mn$ as a set of Lorentz scalars. 
Thus, \eq{nclagr1} violates particle Lorentz symmetry. 
Within a given inertial reference frame, 
$\th^\mn$ may be thought to provide a 4-dimensional 
directionality to spacetime.
This behavior is similar to that of background tensor expectation values 
in spontaneous Lorentz symmetry breaking.\cite{ksp}

The discussion of the previous paragraph may be 
obviously generalized to any noncommutative theory, 
implying that any noncommutative theory 
violates (particle) Lorentz symmetry. 


The recipe described above 
does not directly specify the relationship 
between noncommuative field operators 
and realistic physical variables. 
For example, 
since the fermion field $\hat{\ps}$ in \eq{nclagr1} 
is noncommutative and obeys an 
unconventional gauge transformation law, 
the relationship between its quantum 
and the physical electron is nontrivial. 
However, 
there is a known correspondence between 
noncommutative gauge fields and 
ordinary fields, 
called the Seiberg-Witten map,\cite{sw} 
that yields an ordinary guage theory 
with physical content equivalent to 
the noncommutative gauge theory. 
It is presumably feasible to calculate physical observables 
directly in terms of the noncommutative fields,\cite{csjt,mocioiu} 
though we do not take this approach here.

After applying the Seiberg-Witten map to 
any realistic noncommutative gauge theory, 
the result is an ordinary guage theory 
involving standard-model fields 
that breaks particle Lorentz symmetry 
while preserving observer Lorentz symmetry. 
Meanwhile, a general framework already exists 
that has standard-model gauge symmetries, 
is built from standard-model fields, 
breaks particle Lorentz symmetry, 
and preserves observer Lorentz symmetry.\cite{ck,kp}  
Thus, 
{\it any realistic noncommutative gauge theory 
is physically equivalent to a subset 
of the standard-model extension.}

The correspondence between realistic 
noncommutative theories 
and subsets of the standard-model extension 
allows results from the latter theoretical framework 
to be applied to the former. 
Among the consequences for any realistic 
noncommutative field theory:
\begin{enumerate}
\item Energy and momentum are conserved. 
\item CPT is preserved. 
 However, all other combinations of 
 the discrete symmetries C, P, T may be broken.\cite{sj}
\item The fermionic sector is free of 
 perturbative difficulties with 
 stability and causality.\cite{kl} 
 Accordingly, superluminal information transfer is absent.
\item The conventional spin-statistics relation holds. 
\item There are no difficulties with perturbative unitarity, 
 provided $\th^\mn\th_\mn\ge0$ and
 $\ve^{\mn\al\be}\th_{\mn}\th_{\al\be}=0$.
\end{enumerate}


In the remainder of this work, 
we assume $\th^\mn\th_\mn>0$ 
and $\ve^{\mn\al\be}\th_{\mn}\th_{\al\be}=0$, 
and focus on the noncommutative QED 
described in \eq{nclagr1}. 
Since physical noncommutativity in nature must be small, 
it suffices to consider only effects that are 
leading order in $\th^\mn$. 
In this case, 
the explicit form of the Seiberg-Witten map is known\cite{sw,bichl}:
\bea
\widehat{A}_\mu &=& A_\mu 
 -\half\th^{\al\be}A_\al 
 (\prt_\be A_\mu + F_{\be\mu} ) \nonumber \\
\widehat{\ps} &=& \ps 
 -\half\th^{\al\be}A_\al \prt_\be \ps
\quad .
\label{swmap}
\eea
Combining \eq{moyal}--\eq{swmap}
yields an ordinary quantum field 
theory that is physically equivalent to noncommutative 
QED to leading order in $\th^\mn$. 

We are primarily interested in situations involving 
constant electromagnetic fields 
since our focus is on experiments that satisfy this condition. 
To this end, 
we substitute $F_\mn\rightarrow f_\mn + F_\mn$, 
where $f_\mn$ is a constant background electromagnetic field 
and $F_\mn$ is assumed to be a small dynamical fluctuation.  
We then perform a physically irrelevant rescaling 
of the fields $\ps$ and $A_\mu$
(to preserve conventional normalization of kinetic terms), 
and disregard terms of third order or larger 
in the fluctuations. 
Finally, we redefine the gauge field 
$A_\mu\rightarrow qA_\mu$ to display the charge coupling of 
the physical fermion.

The result of these manipulations is 
the hermitian lagrangian 
\bea
{\cal L} &=& \half i\ol{\ps}\ga^\mu\lrDmu\ps 
 -m\ol{\ps}\ps -\frac{1}{4}F_\mn F^\mn \nonumber \\
&& +\half ic_\mn\ol{\ps}\ga^\mu\lrDnu\ps
 -\frac{1}{4}(k_F)_{\al\be\ga\de}F^{\al\be}F^{\ga\de}
\quad .
\label{nclagr2}
\eea
In this equation,  
$D_\mu\ps = \prt_\mu\ps -iq_{\rm eff}A_\mu\ps$, 
where the charge $q_{\rm eff}$ is a scaled effective value, 
\beq
q_{\rm eff}:= (1+\frac{1}{4}qf^\mn\th_\mn )q
\quad .
\label{defn_qeff}
\eeq
The dimensionless coefficients $c_\mn$ and $k_{F\al\be\ga\de}$ are 
\bea
c_\mn &:=& -\half q{f_\mu}^\la \th_{\la\nu} \quad , 
 \nonumber \\
(k_F)_{\al\be\ga\de} &:=& -q{f_\al}^\la \th_{\la\ga} \et_{\be\de} 
 +\half qf_{\al\ga}\th_{\be\de} -\frac{1}{4}qf_{\al\be}\th_{\ga\de} 
 \nonumber \\
 && -(\al\leftrightarrow\be)-(\ga\leftrightarrow\de) 
 + (\al\be\leftrightarrow\ga\de)
\quad .
\label{ncck}
\eea
The notation here is chosen to resemble  
that of the standard­-model extension 
in its QED limit.\cite{ck} 
Note, however, that the coefficients $c_\mn$ and $(k_F)_{\al\be\ga\de}$ 
now depend on the background electromagnetic field strength, 
so some caution is required in applications. 
Note also that, 
at least to leading order, 
noncommutative effects vanish for 
neutral fermions. 


The photon sector of \eq{nclagr2} has been 
studied elsewhere.\cite{jackiw} 
In the present work, 
we concentrate primarily on the fermion sector. 
In particular, 
we focus on clock-comparison experiments,\cite{ccexpt} 
which place stringent bounds\cite{klane} on 
the parameter $c_\mn$. 

From \eq{nclagr2}, 
we can calculate a hermitian perturbation hamiltonian, 
the expectation value of which 
gives a quadrupole-type energy-level shift 
$\de\sim \tmf\ga m(c_{11}+c_{22}-2c_{33})\sim\tmf\ga m q B\th^{12}$ 
to a fermion of charge $q$ and mass $m$. 
Here $\tmf$ denotes a ratio of Clebsch-Gordan coefficients 
and $\ga$ denotes an expectation value of momentum operators. 
Both $\tmf$ and $\ga$ are zero unless the particle has orbital and total 
angular momentum $l,j\ge 1$. 
In deriving this formula, it has been assumed that the fermion 
is in a constant magnetic field $B$ 
parallel to the laboratory $z$-axis. 

Through $\th^{12}=\th(\hat{x},\hat{y})$, 
which is given in terms of a laboratory basis $(\hat{x},\hat{y},\hat{z})$, 
the energy shift $\de$ varies with time 
as Earth rotates. 
To explicitly display this variation, 
we re-express $\de$ with respect to a nonrotating frame $(\hat{X},\hat{Y},\hat{Z})$: 
\beq
\de = E_0 + E_{1X}\cos\Om t + E_{1Y}\sin\Om t
\quad .
\label{delta}
\eeq
In this expression, 
$E_0$ is an irrelevant constant 
and $(E_{1X},E_{1Y})\sim(\th^{YZ},\th^{ZX})$ give the amplitude of 
the variation of $\de$ with sidereal frequency $\Om$. 
Contrast this with the situation when $c_\mn$ is 
independent of $B$, 
in which case $\de$ also has variation at frequency $2\Om$. 


We can apply these results to recent clock­-comparison 
tests.\cite{ccexpt}
Most are sensitive only to Lorentz-violating effects 
in the neutron, 
and so are insensitive to noncommutative-geometry effects. 
However, 
two experiments contain sensitivity to charged particles, 
and can be used to bound the scale of noncommutative geometry 
in nature. 
 
The experiment of Berglund et al.\ 
bounds sidereal variations in 
certain $\cs$ and $\hg$ transitions to about 100 nHz. 
However, the weak magnetic field $B\sim 5$ mG 
used in the experiment 
leads to a relative suppression of noncommutative effects, 
leaving a bound 
$\abs{\th^{YZ}},\abs{\th^{ZX}}\lsim (10\ {\rm GeV})^{-2}$. 

The experiment of Prestage et al.\ 
bounds sidereal variations in 
certain $\ber$ transitions to about 100 $\mu$Hz. 
In contrast to Berglund et al., 
this experiment used a rather strong applied magnetic 
field $B\sim 1\ T$. 
The resulting bound on noncommutativity is 
\beq
\abs{\th^{YZ}},\abs{\th^{ZX}}\lsim\ (10\ {\rm TeV})^{-2} 
\quad .
\label{bound}
\eeq

The noncommutative parameter $\th^\mn$ 
has been bounded in other low-energy experiments. 
For example, 
a bound several orders of magnitude weaker than \eq{bound} 
arises from study of the Lamb shift.\cite{csjt} 
Elsewhere, an analysis involving anomalous spin couplings 
and coherent nuclear effects\cite{mocioiu} leads to a 
speculative bound some 20 order of magnitude 
stronger than \eq{bound}. 


\section*{References}
\begin{thebibliography}{99}

\bibitem{cds} A.\ Connes, M.\ Douglas and A.\ Schwarz, JHEP {\bf 02}, 
003 (1998). For recent reviews, see N.A.\ Nekrasov, 
hep-th/0011095; A.\ Konechny and A.\ Schwarz, hep-th/0012145; 
J.A.\ Harvey, hep-th/0102076. 

\bibitem{cpt99} Some reviews are in V.A.\ Kosteleck\'y, ed., 
{\it CPT and Lorentz Symmetry}, World Scientific, Singapore, 1999 
and the current proceedings. 

\bibitem{chklo} S.\ Carroll, et al., Phys.\ Rev.\ Lett.\ {\bf 87} 141601 (2001).

\bibitem{haya} M.\ Hayakawa, Phys.\ Lett.\ B {\bf 478}, 394 (2000). 

\bibitem{ck} D.\ Colladay and V.A.\ Kosteleck\'y, Phys.\ Rev.\ D {\bf 55}, 6760 
(1997); Phys.\ Rev.\ D {\bf 58}, 116002 (1998). 

\bibitem{ksp} 
V.A.\ Kosteleck\'y and S.\ Samuel, Phys.\ Rev.\ D {\bf 39}, 683 
(1989); ibid.\ {\bf 40}, 1886 (1989); Phys.\ Rev.\ Lett.\ {\bf 63}, 224 
(1989); V.A.\ Kosteleck\'y and R.\ Potting, Nucl.\ Phys.\ B 
{\bf 359}, 545 (1991); Phys.\ Lett.\ B {\bf 381}, 89 (1996). 

\bibitem{sw} N.\ Seiberg and E.\ Witten, JHEP {\bf 09}, 032 (1999). 

\bibitem{kp} V.A.\ Kosteleck\'y and R.\ Potting, Phys.\ Rev.\ D {\bf 51}, 3923 
(1995). 

\bibitem{sj} M.M.\ Sheikh-Jabbari, Phys.\ Rev.\ Lett.\ {\bf 84}, 5265 (2000). 

\bibitem{kl} V.A.\ Kosteleck\'y and R.\ Lehnert, Phys.\ Rev.\ D {\bf 63}, 
065008 (2001). 

\bibitem{bichl} A.A.\ Bichl et al., hep-th/0102103. 

\bibitem{jackiw} R.\ Jackiw, these proceedings; 
 Z.\ Guralnik et al., hep-th/0106044.

\bibitem{ccexpt} J.D.\ Prestage et al., Phys.\ Rev.\ Lett.\ {\bf 54}, 2387 (1985); 
S.K.\ Lamoreaux et al., Phys.\ Rev.\ Lett.\ {\bf 57}, 3125 (1986); 
Phys.\ Rev.\ A {\bf 39}, 1082 (1989); T.E.\ Chupp et al., Phys.\ 
Rev.\ Lett.\ {\bf 63}, 1541 (1989); C.J.\ Berglund et al., Phys.\ 
Rev.\ Lett.\ {\bf 75}, 1879 (1995); D.\ Bear et al., Phys.\ Rev.\ 
Lett.\ {\bf 85}, 5038 (2000). 

\bibitem{klane} V.A.\ Kosteleck\'y and C.D.\ Lane, Phys.\ Rev.\ D {\bf 60}, 116010 
(1999); J.\ Math.\ Phys.\ {\bf 40}, 6245 (1999). 

\bibitem{csjt} M.\ Chaichian, M.M.\ Sheikh-Jabbari, and A.\ Tureanu, 
Phys.\ Rev.\ Lett.\ {\bf 86}, 2716 (2001). 

\bibitem{mocioiu} I.\ Mocioiu, these proceedings. 



\end{thebibliography}
\end{document}